\documentclass[twocolumn,showpacs,amsmath,amssymb, A4,superscriptaddress]{revtex4-1}

\usepackage{color}

\usepackage{colordvi}
\usepackage{amssymb}
\usepackage{amsmath}
\usepackage{epsf}
\usepackage{mathrsfs}
\usepackage{graphicx}
\usepackage{appendix}

\def\nn{\nonumber}
\def\be{\begin{equation}}
\def\ee{\end{equation}}
\def\bea{\begin{eqnarray}}
\def\eea{\end{eqnarray}}
\begin{document}

\title{\textbf{Two-leg ladder Bose Hubbard models with staggered fluxes}}

\author{Rashi Sachdeva}
\email{rashi.sachdeva@oist.jp}
\affiliation{Quantum Systems Unit, Okinawa Institute of Science and Technology Graduate University, Okinawa 904-0495, Japan}

\author{$\!\!^{,}\,^{\dagger}\,\,\,$Friederike Metz}
\thanks{These two authors contributed equally.}
\affiliation{Quantum Systems Unit, Okinawa Institute of Science and Technology Graduate University, Okinawa 904-0495, Japan}
\thanks{These two authors contributed equally.}

\author{Manpreet Singh}
\affiliation{Department of Physics, Indian Institute of Technology, Guwahati-781039, Assam, India }

\author{Tapan Mishra}
\affiliation{Department of Physics, Indian Institute of Technology, Guwahati-781039, Assam, India }

\author{Thomas Busch}
\affiliation{Quantum Systems Unit, Okinawa Institute of Science and Technology Graduate University, Okinawa 904-0495, Japan}

\date{\today}

\begin{abstract}

We investigate the ground state properties of ultracold atoms trapped in a two-leg ladder potential in the presence of an artificial magnetic field in a staggered configuration. We focus on the strongly interacting regime and use the Landau theory of phase transitions and a mean field Gutzwiller variational method to identify the stable superfluid phases and their boundaries with the Mott-insulator regime as a function of magnetic flux. In addition, we calculate  the local and chiral currents of these superfluid phases, which show a staggered vortex anti-vortex configuration. The analytical results are confirmed by numerical simulations using a cluster mean-field theory approach.
 \end{abstract}

\maketitle

\section{Introduction}
 Ultracold bosonic atoms in optical lattices offer a unique platform to study models for periodic many body physics in a clean and highly controllable setting.  A wide range of flexible geometries to trap neutral atoms can be created by overlapping and interfering laser beams and interactions can be controlled via external magnetic fields or by choosing different atomic species. While the field was initially enthused by the prediction and realisation of the paradigmatic superfluid to Mott-insulator transition in square lattices \cite{jaksch,bloch}, many different situations have been investigated since  then \cite{lewenstein_book, bloch_review}. 
  
Recent progress in creating artificial gauge fields for ultracold atoms in discrete~\cite{magfield_OL} as well as continuum systems~\cite{magfield_cont}  has opened up many new avenues for the study of quantum phase transitions in the presence of magnetic fields.  These fields are called artificial, as due to the charge neutrality of the atoms no Lorentz force exists and therefore real magnetic fields do not directly effect  the center-of-mass variable. 

The simplest way to mimic the effects of magnetic fields on charged systems in neutral atoms is by rotation \cite{rot_exp}, which probes superfluidity in the same way  magnetic fields probe superconductivity. Furthermore, very high synthetic magnetic fields have been shown to be realizable using atoms in optical lattices, where the atomic motion and the internal degrees of freedom can be coupled by laser assisted tunneling  \cite{bloch_PRL2011}. This has lead to the successful implementation of uniform as well as staggered flux distributions in the strong field regime \cite{bloch_PRL2011, sengstock_PRL2012} and has enabled the realization of 2D topological states with finite Chern numbers \cite{chern_bloch15, chern_bloch18}.

Theoretically,  the presence of artificial magnetic fields can be included into the Bose Hubbard model by using complex tunnel couplings \cite{bhm_agf}. The main effect of these can be observed even in the absence of interactions and the single particle spectrum for bosons in a periodic potential in the presence of a strong magnetic field forms a self-similar structure known as the Hofstadter butterfly~\cite{hofstadter}. As the effective magnetic fields created in optical lattices can be much larger than what is possible in solid-state systems, these techniques bring the study of a wide range of Hamiltonians into reach that are inaccessible in condensed matter physics. 

Besides the realization of magnetic fields in extended 2D lattice systems, the effects of artificial magnetic fields were also studied in bosonic ladder geometries, where chiral currents and vortex and Meissner phases were predicted and observed  \cite{georges_NJP2014, tokuno_PRA2015, natu_PRA2015, giamarchi_PRB2001,oktel_PRA2015, mueller_PRA2014, rashi_PRA2017, bloch_ladder}. While ladder systems can be seen as the smallest possible lattice structure, they possess additional and unique properties, for example due to the absence of the requirement that the magnetic fields have to have rational values ~\cite{giamarchi_PRB2001, oktel_PRA2015, mueller_PRA2014, rashi_PRA2017}. Furthermore, even though the above-mentioned Meissner and vortex phases can already be observed for non-interacting systems, interacting bosonic ladder systems with uniform flux also support various spontaneously symmetry broken phases and chiral Mott insulator states \cite{arya_PRA}.  

Similar to the case of uniform fluxes, staggered fluxes  \cite{staggered_1,staggered_2, staggered_3, staggered_4}  can drive quantum phase transitions in the two-leg Bose Hubbard ladder systems and can enlarge the range of physical effects that can be investigated. Here we study the example of a single-component BEC trapped in such a geometry in the presence of a periodically flipped artificial magnetic field. We find that the presence of the staggered flux gives rise to two superfluid phases with a staggered vortex anti-vortex configuration, which are distinct from the usual superfluid phases obtained in the Bose Hubbard model~\cite{jaksch}. 

The manuscript is organized as follows. In Section \ref{Sec: bhm} we introduce the Bose-Hubbard model (BHM) with a two-leg ladder geometry in the presence of an artificial magnetic field with a staggered configuration. In Section \ref{Sec: singleparticle} we review the properties of its single particle spectrum and in Section \ref{Sec: Landau} we present calculations in the strong coupling regime to determine the complete phase diagram. We also show the presence of distinct superfluid phases using Landau theory. In Section \ref{Sec: Gutzwiller} we present our analytical calculations to determine the phase boundaries using the variational Gutzwiller approach and in Section \ref{Sec: ClusterMF} these are complemented by the numerical calculations performed using the cluster mean field theory approach. Finally, in Section \ref{Sec: Summary} we present a summary and outlook of the work done. 

\section{Model}
\label{Sec: bhm}

\begin{figure}[tb]
	\includegraphics[width=\columnwidth]{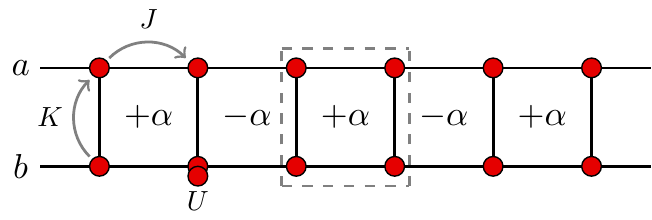}
	\caption{(Color online) Schematic of the two-leg ladder Bose Hubbard model with staggered flux $\alpha$ in neighboring plaquettes. The dashed box indicates the single unit cell used for the analytic and the cluster mean field calculations. The red dots represent the bosonic atoms on lattice sites. }
	\label{fig:schematic}
\end{figure}

The Hamiltonian describing bosons in a two-leg ladder geometry in the presence of a staggered magnetic flux of magnitude $\alpha$ can be written as
\begin{align}
  H=&-J\sum_{j}\left(e^{(-1)^{j}\frac{i\alpha}{2}}a_j^\dagger a_{j+1}+e^{(-1)^{j+1}\frac{i\alpha}{2}}b_j^\dagger b_{j+1}+ h.c.\right)\nonumber\\
      &- K\sum_{j}(a_j^\dagger b_j+ h.c.)+{U \over 2} \sum_{j,p} n_j^p(n_j^p-1)\nonumber\\
      &-\mu\sum_{j,p}n_j^p, 
      \label{eq:eq1_model}
\end{align}
where the $p_j (p_j^\dagger)$ are the bosonic annihilation (creation) operators at site $j$ of leg $p~(=a,b)$,
$n_j^p$ is the number operator at site $j$ of leg $p$, $\alpha$ is the absolute value of the magnetic flux and $\mu$ is the chemical potential. 
The intra- and inter-leg hopping amplitudes are described by $J$ and $K$ respectively, and the on-site interaction energy between two atoms is given by $U$ (see Fig.~\ref{fig:schematic}). The ratios $J/U$ and $K/U$ can be changed in an experiment by tuning the optical lattice laser intensities along each leg and by varying the separation between the legs, respectively. We assume up-down symmetry for the ladder, which implies that the chemical potential $\mu$ and the onsite interactions $U$ are identical for each of the two legs. It is worth noting that within the local density approximation, the results from this model can also be applied to experimental systems which have an additional harmonic trapping potential. 

The phase $\alpha$ appearing in the hopping terms is given by $\alpha=(e/\hbar)\int_{r_j}^{r_k}d\mathbf{r}\cdot\mathbf{A(r)}$, where $\mathbf{A(r)}$ is the vector potential that gives rise to the magnetic field $\mathbf{B}=\nabla \times\mathbf{A}$ and $r_j$ and $r_k$ are the positions of the lattice sites $j$ and $k$. If an atom tunnels around a plaquette, the total phase accumulated by the wavefunction is called the gauge flux, which is a gauge invariant quantity.  Specifically, we choose a Landau gauge for which the hopping in the rung direction has no gauge field while hopping along the legs imparts a phase that alternates from one plaquette to the next, leading to the required staggered flux. The physical properties of the Hamiltonian (\ref{eq:eq1_model}), including the energy spectrum, response functions etc., are of course gauge invariant and only depend on the total flux going through a plaquette.

\section{Single particle spectrum}
\label{Sec: singleparticle}

\begin{figure*}[!htbp]
\includegraphics[width=1.67\columnwidth]{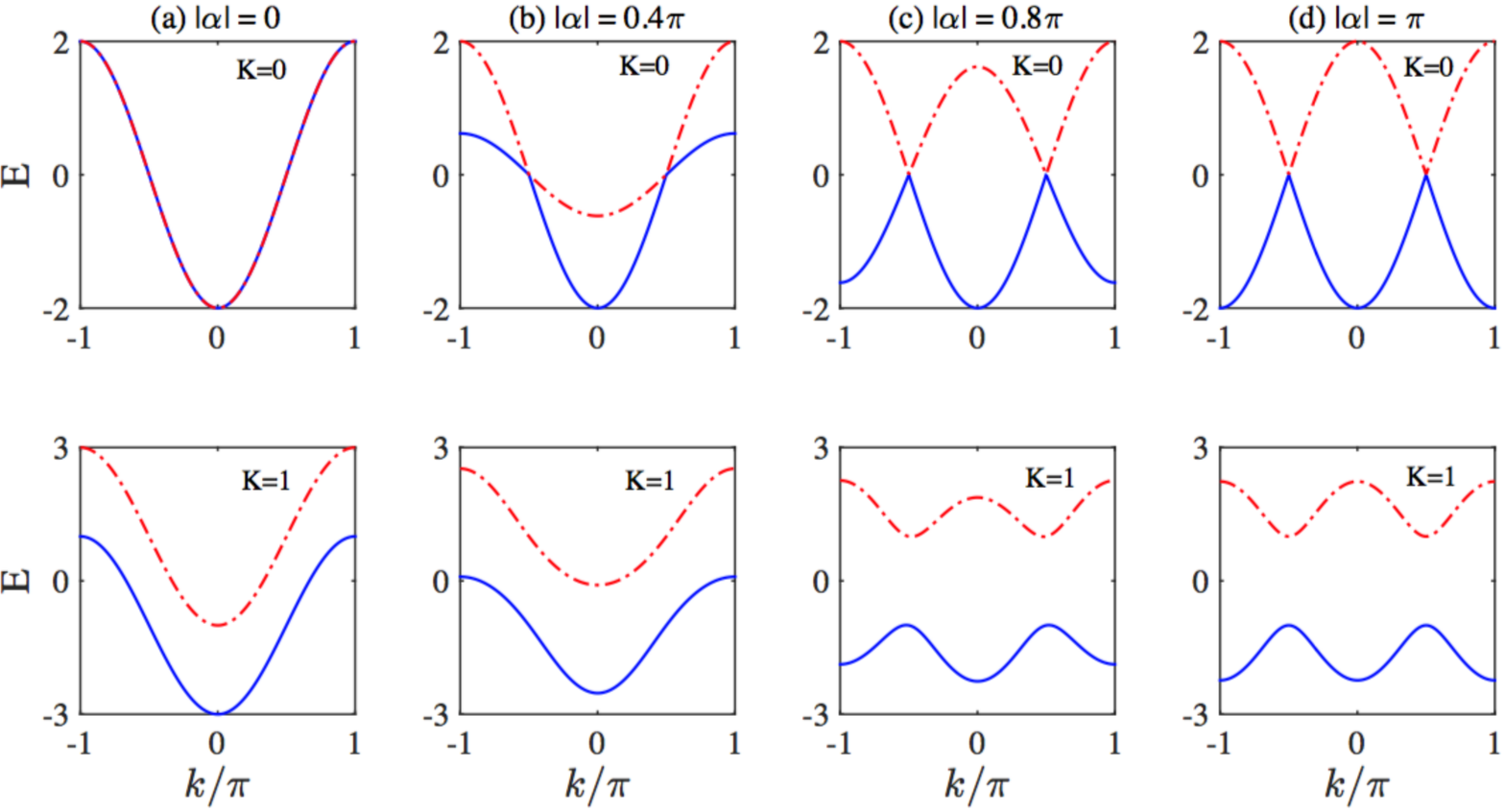}
\caption{(Color online) Single-particle spectrum of the two-leg ladder system for different absolute magnetic flux strengths, and for $K=0$ (top row) and $K=1$ (bottom row). The intra-leg hopping ratio is set to $J=1$.}
\label{Fig:singleparticle_dispersion}
\end{figure*}

We first determine the structure of the single particle energy spectrum as a function of the magnetic flux values. For this we set $U=0$ and write the Hamiltonian in momentum space in terms of the Fourier components of the field operators $a_j$ and $b_j$. The energy eigenvalues can then be determined by simple diagonalization, and we show the spectrum as a function of momentum $k$ in Fig.~\ref{Fig:singleparticle_dispersion}, for different absolute values of the magnetic flux $\alpha$. 

For zero flux and no rung hopping ($K=0$) the single particle spectrum has only one doubly-degenerate band, since the two legs of the ladder are decoupled.  For finite rung coupling ($K=1$) this degeneracy is lifted and a two-band structure appears, which has the expected $2\pi$ periodicity (see Fig.~\ref{Fig:singleparticle_dispersion}(a)). In the presence of a finite staggered flux the lowest band continues to have a non-degenerate minimum at $k=0$ (see Figs.~\ref{Fig:singleparticle_dispersion}(b) and (c)) and increasing the rung coupling $K$ leads to an increase in the band gap between the upper and lower bands. 
Since the system now possesses a finite flux, condensing into the minimum leads to a superfluid with a unique current pattern which is further discussed in Sec.~\ref{Sec: Landau}. Upon increasing the staggered flux further, the lowest band starts developing additional minima at $k=\pm \pi$ (see Fig.~\ref{Fig:singleparticle_dispersion}(c)), which eventually become degenerate with the minimum at $k=0$ for $\alpha=\pi$ (see Fig.~\ref{Fig:singleparticle_dispersion}(d)). This limit is known as the \textit{fully frustrated} case for the Bose Hubbard model and it corresponds to half a flux quantum per plaquette \cite{arya_PRA}.

The occurence of degenerate minima at $k/\pi=0$ and $k/\pi=\pm 1$ can influence the stability and properties of the phases in different regimes. While for the Mott-insulating regime the qualitative nature of the phase remains unaffected, the properties of the superfluid states get substantially changed due to the staggered flux. We discuss this situation in detail in the next section.

\section{Superfluid Mott-insulator transition: Landau theory of phase transitions} 
\label{Sec: Landau}

In this section we discuss the results obtained for strong coupling regime and determine the complete phase diagram at zero temperature. For the Bose Hubbard model with no flux, the zero-temperature phase diagram comprises a superfluid (SF) phase and a Mott insulator (MI) phase, which are separated by a second-order phase transition, driven by quantum fluctuations \cite{Fisher89}. When one crosses the phase boundary from MI into the SF phase, the $U(1)$ gauge symmetry is spontaneously broken, which gives rise to a finite SF-order parameter. Since the form of this order parameter depends on system parameters, one can expect that the presence of  a finite staggered flux leads to different and distinctly broken-symmetry SF phases. In the following we will use the Landau theory of phase transitions and introduce a plaquette order parameter, which identifies the various SF phases. Determining the values of $U/J$ at which the SF order parameter vanishes allows us to obtain the phase boundaries within the full phase diagram as a function of the magnetic flux $\alpha$. 

The basic plaquette in our system consists of four sites, indicated by the dashed lines in Fig.~\ref{fig:schematic}. The different superfluid phases will be characterised by introducing the plaquette order parameter $\Psi=(\psi_1,\chi_1,\chi_2,\psi_2)$, where $\psi_{i}=\langle a_{i}\rangle$ and $\chi_{i}=\langle b_{i}\rangle$ stand for site order parameters for legs $a$ and $b$, respectively. 
In the mean-field limit we can decouple the sites of the unit cell by \cite{decoupling_EPL}
\begin{align}
a_j^{\dag}a_k&\approx \psi_j^{\ast}a_k+a_j^{\dag}\psi_k-\psi_j^{\ast}\psi_k, \nn\\
b_j^{\dag}b_k& \approx \chi_j^{\ast}b_k+b_j^{\dag}\chi_k-\chi_j^{\ast}\chi_k,\nn\\
a_j^{\dag}b_j& \approx \psi_j^{\ast}b_j+a_j^{\dag}\chi_j-\psi_j^{\ast}\chi_j,  
\end{align}
where $j,k~\in \{1,2\}$. Hence, we can write the mean field Hamiltonian in the grand canonical ensemble in the form 
\bea H=H_{0}^{\text{MF}}+H_{1}^{\text{MF}}\nn,\eea
where
\begin{align}
 H_{0}^{\text{MF}}=& {U \over 2} \sum_{j=1,2} (n_j^a(n_j^a-1)+n_j^b(n_j^b-1))\nn\\
 & -\mu \sum_{j=1,2} (n_j^a+n_j^b)+K\sum_{j=1,2} (\psi_j^{\ast}\chi_{j}+\chi_{j}^{\ast}\psi_j)\ \nn\\
 & +J\sum_{j=1} (e^{-i\alpha}\psi_j^{\ast}\psi_{j+1}+e^{i\alpha}\chi_j^{\ast}\chi_{j+1}+h.c.)\nn\\
 & +J\sum_{j=2} (e^{i\alpha}\psi_j^{\ast}\psi_{j+1}+e^{-i\alpha}\chi_j^{\ast}\chi_{j+1}+h.c.),\\
 H_{1}^{\text{MF}}=& -J\sum_{j=1}\big( e^{-i\alpha}\psi_j^{\ast}a_{j+1}+e^{-i\alpha}\psi_{j+1}a_{j}^{\dag}+e^{i\alpha} \chi_{j}^{\ast}b_{j+1}\nn\\
 & +e^{i\alpha}\chi_{j+1}b_{j}^{\dag}+h.c\big)-J\sum_{j=2}\big( e^{i\alpha}\psi_j^{\ast}a_{j+1}\nn\\
 & +e^{i\alpha}\psi_{j+1}a_{j}^{\dag} +e^{-i\alpha} \chi_{j}^{\ast}b_{j+1} +e^{-i\alpha}\chi_{j+1}b_{j}^{\dag}+h.c.\big)\nn\\
 & -K\sum_{j=1,2}\left(\psi_j^{\ast}b_j+a_j^{\dag}\chi_j+h.c.\right).
 \end{align}
Since we concentrate on the strong-coupling regime, our expansion will treat  $H_{1}^{\text{MF}}$ as a perturbation.
Calculating the ground state energy, $E[\psi]$, for the four site plaquette up to second order with respect to the perturbation $H_{1}^{\text{MF}}$ then gives
\begin{equation}
 E[\Psi]= 2Un(n-1)-4\mu n+\sum_{\nu,\nu'}\Psi_{\nu}^{\ast}M_{\nu,\nu'}\Psi_{\nu'} ,
 \end{equation}
 where $n$ is the filling fraction and $M_{\nu,\nu'}$ are the matrix elements of the $4\times4$ Hermitian matrix $M$ which is given by 
\begin{widetext}
\[
  M=
  \left[ {\begin{array}{cccc} 
   E_{0}(K^2+4J^2) & K  & 4KJE_0~\text{cos}(\frac{\alpha}{2}) & 2Je^{-i\alpha/2}\\
     K & E_{0}(K^2+4J^2) & 2Je^{i\alpha/2} &  4KJE_0~\text{cos}(\frac{\alpha}{2}) \\
     4KJE_0~\text{cos}(\frac{\alpha}{2})  &2Je^{-i\alpha/2}  & E_{0}(K^2+4J^2) &  K  \\
     2Je^{i\alpha/2}  & 4KJE_0~\text{cos}(\frac{\alpha}{2}) & K & E_{0}(K^2+4J^2) \\
    \end{array} } \right],
\]
with
\bea E_{0}(n,U,\mu)=\bigg[\frac{n}{U(n-1)-\mu}+\frac{n+1}{\mu-Un}\bigg].\eea
In standard Landau theory, the free energy is expanded with respect to a scalar order parameter and the 
phase transition boundary is determined by demanding that the second-order expansion coefficient should vanish. In our case, the second order phase transitions between the different SF and MI phases therefore occur when the eigenvalues of $M$ are zero. The matrix has four eigenvalues and eigenvectors given by 
\begin{align}
\epsilon_1&=E_0(4J^2+K^2+4JK\text{cos}(\alpha/2))+\sqrt{4J^2+K^2+4JK\text{cos}(\alpha/2)},\\
\epsilon_2&=E_0(4J^2+K^2-4JK\text{cos}(\alpha/2))+\sqrt{4J^2+K^2-4JK\text{cos}(\alpha/2)},\\
\epsilon_3&=E_0(4J^2+K^2+4JK\text{cos}(\alpha/2))-\sqrt{4J^2+K^2+4JK\text{cos}(\alpha/2)},\\
\epsilon_4&=E_0(4J^2+K^2-4JK\text{cos}(\alpha/2))-\sqrt{4J^2+K^2-4JK\text{cos}(\alpha/2)},\\
\end{align}

\begin{align}
\Psi_{1}&=\left( \frac{K+ 2Je^{i\alpha/2}}{|K+ 2Je^{i\alpha/2}|},1,\frac{K+ 2Je^{i\alpha/2}}{|K+ 2Je^{i\alpha/2}|},1\right)&&\hspace*{-100pt}=\left( e^{i\theta_1},1,e^{i\theta_1},1\right)\label{SF1},\\ 
\Psi_{2}&=\left(-\frac{K- 2Je^{i\alpha/2}}{|K- 2Je^{i\alpha/2}|},-1, \frac{K- 2Je^{i\alpha/2}}{|K- 2Je^{i\alpha/2}|},1\right)&&\hspace*{-100pt}=\left(-e^{i\theta_2},-1, e^{i\theta_2},1\right)\label{SF2},\\
\Psi_{3}&=\left(-\frac{K+ 2Je^{i\alpha/2}}{|K+ 2Je^{i\alpha/2}|},1,-\frac{K+ 2Je^{i\alpha/2}}{|K+ 2Je^{i\alpha}|},1\right)&&\hspace*{-100pt}=\left(-e^{i\theta_1},1,-e^{i\theta_1},1\right),\\
\Psi_{4}&=\left(\frac{K- 2Je^{i\alpha/2}}{|K- 2Je^{i\alpha/2}|}, -1,-\frac{K- 2Je^{i\alpha/2}}{|K- 2Je^{i\alpha}|},1\right)&&\hspace*{-100pt}=\left(e^{i\theta_2}, -1,-e^{i\theta_2},1\right),
\end{align}
where $\theta_1=\text{tan}^{-1}(\frac{2J\text{sin}(\alpha/2)}{K+2J\text{cos}(\alpha/2)})$ and $\theta_2=\text{tan}^{-1}(-\frac{2J\text{sin}(\alpha/2)}{K-2J\text{cos}(\alpha/2)})$. These four eigenvectors describe all possible SF phases.
\end{widetext}

\subsection{Interpretation of the superfluid phases}
One can see that all the four eigenvectors depend explicitly on the flux $\alpha$ and are complex for certain values of $\alpha$. However, careful examination of the above four eigenvalues shows that only two eigenvectors, $\Psi_1$ and $\Psi_2$, correspond to stable superfluid phases for repulsive onsite interactions for different regimes of magnetic flux $\alpha$.

In the following we label these two distinct SF phases as superfluid 1 (SF-1) and superfluid 2 (SF-2). They are characterized by circulating gauge invariant currents around the plaquettes, which are arranged in a staggered pattern along the ladder and can be viewed as a sequence of vortices and anti-vortices, as shown by the chiral current calculations in section \ref{currents}. Both possess a spatially uniform boson density, but the sign of leg/rung currents correspond to two distinct current order patterns which are related to one another by time reversal or by a unit translation.  This is consistent with the results known for the fully frustrated case with $\alpha=\pi$ flux per plaquette, where Hartree theory indicates the presence of the same two superfluid phases \cite{arya_PRA}.  Since the Hamiltonian is both translationally and time-reversal invariant, the emergence of these staggered flux states is a result of the breaking of these symmetries and we detail the calculation for staggered gauge-invariant currents for the SF-1 and the SF-2 phase in Section \ref{currents}.

The phases of the order parameters at each lattice site are given by $\Phi_\text{SF-1} = \left(\theta_1,0, \theta_1,0\right) $ for SF-1 and $\Phi_\text{SF-2} = \left(\theta_2+\pi,\pi, \theta_2,0\right)$ for SF-2. In the fully frustrated case, which is the point where the system switches between being in SF-1 and SF-2, the phase around the plaquette for both superfluid states becomes equal and opposite, manifesting the opposite circulation of currents in each state.  At this particular value of the magnetic flux, the energy eigenvalues of both superfluid states become degenerate as well, and while for $\alpha<\pi$ the SF-1 phase had the lower energy, beyond $\alpha=\pi$ the SF-2 become energetically more favourable. 
This transition from the SF-1 to the SF-2 phase therefore corresponds to a reversal of the direction of circulation.

\subsection{Phase diagram}
\begin{figure}[tb]
\includegraphics[width=\columnwidth]{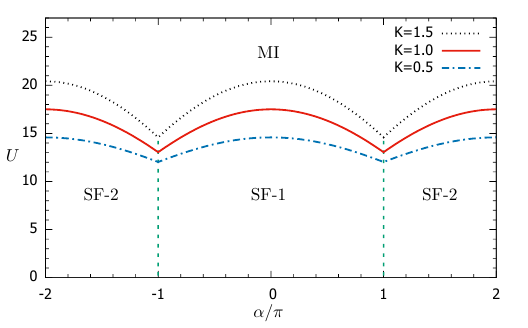}
\caption{(Color online) Phase diagram for the two-leg ladder Bose Hubbard model in the presence of a staggered flux of magnitude $\alpha$ for unit filling factor using Landau theory. The solid (red) curve marks the boundary between the Mott-insulator and the different superfluid phases for $K=J=1.0$. The region below the solid (red) curve comprises of two types of superfluids, SF-1 and SF-2 (see text for details) which are separated by green dashed lines. The dashed (blue) lines and dotted (black) lines mark the phase boundaries for $J=1$ and $K=0.5$ and $1.5$, respectively.}
\label{Fig:phase_ana}
\end{figure}

The boundary between the MI and SF phases can be found as a function of $\alpha$ by determining the zeros of the respective eigenvalues and we show the full phase diagram in Fig.~\ref{Fig:phase_ana}. The zero crossings exist in the range $-\pi < \alpha<\pi$ for SF-1, and in the ranges $-3\pi < \alpha<-\pi$ and $\pi < \alpha<3\pi$ for SF-2, implying a $2\pi$ periodicity for both the superfluid phases.
As noted above, for values of $\alpha$ beyond $\pm\pi$, the SF-1 undergoes a transition to the SF-2, which at this point becomes energetically favourable($\epsilon_2 < \epsilon_1$). The critical point of transition from SF to MI phase for $\alpha=0$ agrees with the known mean field results \cite{oktel_PRA2015}. It is also worth nothing that at $\alpha=\pi$ and $-\pi$, for a gauge choice where the phase $\alpha$ is only along one of the legs, the Hamiltonian is real and therefore time-reversal invariant. 

The phase diagram as a function of different values of the hopping amplitude $K$ with fixed $J$ is shown in Fig.~\ref{Fig:phase_ana}. For $K<1$, the hopping along the rung of the ladder is reduced, and hence the transition to the Mott-insulating state can be achieved at lower values of $U$. Similarly, for $K>1$ the overall hopping is larger compared to the situation with $K=1$ and the transition to the Mott-insulating phase requires a higher value of the onsite interaction $U$. This suggests that one can tune the phase transition boundary by simply changing the relative hopping amplitudes for any value of flux $\alpha$. 

\section{Variational Mean field Gutzwiller approach for phase boundaries} \label{Sec: Gutzwiller}

In the following we will explore the transition from the Mott-insulator to the above mentioned distinct superfluid phases as a function of $J$, $U$, $\mu$ and $\alpha$. For this we scale the Hamiltonian in Eq.~\eqref{eq:eq1_model}  by setting $K=1$ and assume that the wavefunction for the perfect Mott-insulating phase is localized with an equal number of particles $n_0$ at each site. The phase boundary between the incompressible MI phase and the compressible SF phases can then be  analytically determined by calculating the energy for particle-hole-type excitations using a reduced-basis variational ansatz for the Gutzwiller wave function. 

For this we assume that the total wavefunction is the product of two individual ladder wavefunctions,  $|\Psi\rangle=\Pi_j |G \rangle_{a_j}  |G \rangle_{b_j} $, where $a$ and $b$ label the legs of the ladder and $j$ the individual sites along a leg. In the strongly interacting regime, we work very near to the phase boundary, which implies that only Fock states close to the MI one are populated. Hence we can write a Gutzwiller ansatz for the local sites as
\begin{align}
	 |G\rangle_{a_j}  &= f_{n_0-1}^{a_j}|n_0-1\rangle+f_{n_0}^{a_j}|n_0\rangle+f_{n_0+1}^{a_j}|n_0+1\rangle \nn\\
	|G\rangle_{b_j}  &= f_{n_0-1}^{b_j}|n_0-1\rangle+f_{n_0}^{b_j}|n_0\rangle+f_{n_0+1}^{b_j}|n_0+1\rangle. 
\end{align}

\begin{figure}[tb]
\centering
\includegraphics[width=\columnwidth]{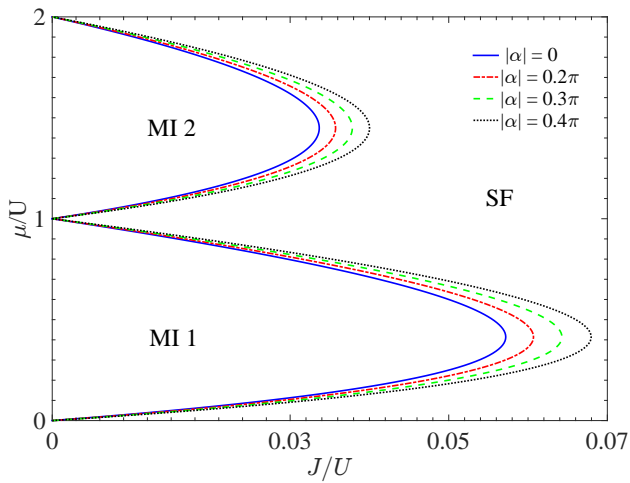}
\caption{(Color online) Phase diagram of the Bose Hubbard model for the two leg ladder for different absolute values of staggered magnetic flux $\alpha$, for K = 1 and U = 1, calculated using a variational mean field approach. The MI phases are indicated with their average occupancy per site, and SF indicated in the plot can be SF-1 for $-\pi<\alpha<\pi$ and SF-2 for the regime $-3\pi<\alpha<-\pi$ and $\pi<\alpha<3\pi$.}
\label{Fig:gutzwiller}
\end{figure}

We parameterise the amplitudes as  \cite{Gutzwiller}
\begin{widetext}
\begin{align} 
(f_{n_0-1}^{a_j}, f_{n_0}^{a_j}, f_{n_0+1}^{a_j})&=(e^{-i\theta_{j}}\Delta_{a_j},\sqrt{1-\Delta_{a_j}^2-\Delta_{a_j} ^{'2}},e^{i\theta_{j}}\Delta_{a_j}^{'}),\\
 (f_{n_0-1}^{b_j}, f_{n_0}^{b_j}, f_{n_0+1}^{b_j})&=(e^{-i\theta_{j}}\Delta_{b_j},\sqrt{1-\Delta_{b_j}^2-\Delta_{b_j} ^{'2}},e^{i\theta_{j}}\Delta_{b_j}^{'}), 
 \label{coefficients}
 \end{align}
 \end{widetext}
with complex variational parameters $\Delta_{a_j}, \Delta_{a_j} ^{'},\Delta_{b_j}, \Delta_{b_j}^{'}\ll1$ to ensure the normalisation condition of states $ |G \rangle_{a_j}$ and $|G \rangle_{b_j}$. Minimizing the energy functional with respect to the variational parameters $\Delta_{a_j}, \Delta_{a_j} ^{'},\Delta_{b_j}, \Delta_{b_j}^{'}$ and $\theta_{j}$, gives the boundary between the MI and SF phase for any value of $\mu$, $U$, and $\alpha$. The dependence on the value of magnetic flux is implicit in the largest eigenvalue of the single particle Hamiltonian and the Mott-insulator/superfluid phase boundaries are shown as a function of the magnetic flux $\alpha/\pi$ and interaction strength $U$ in Fig.~\ref{Fig:gutzwiller}.

It can be seen that a higher magnetic flux enlarges the regions where the Mott-insulator phase appear by shifting the critical point or tip of the lobe for the phase transition to higher values. This enlargement of the insulating phase is expected since the effect of the magnetic field is to localize the single particle dynamics even for non-interacting systems, thus making the transition to an insulating phases easier.
 
Let us stress that these results are exact within mean field theory. The shape of the MI lobe is concave and independent of the dimensionality, since in our mean field calculations the dimensionality enters only through a prefactor. Since fluctuations are known to be particularly important in lower dimensions, one cannot expect the mean field theory to be quantitatively accurate for quasi one-dimensional systems. Hence, the results from the above analysis carry only qualitative importance, and provide a general idea of how the phase boundaries are affected by the presence of magnetic flux. In particular, they can be expected to work only for small hopping strengths when correlations are weak. To complete our study, we present in the following numerical calculations for the phase diagram and the chiral currents.

\section{Numerical Results} \label{Sec: ClusterMF}
In the following we analyze the model given in Eq.(\ref{eq:eq1_model}) numerically using a self-consistent cluster mean-field theory (CMFT) approach. For this a cluster of sites is considered as a unit cell of the system which is then decoupled from all other clusters using 
the mean-field decoupling approximation. 
For any two adjacent sites ($i,j$) which belong to 
different clusters we therefore write
\begin{equation}
 a_i^\dagger a_j \approx \phi_i^* a_j + a_i^\dagger \phi_j - \phi_i^* \phi_j,
\end{equation}
where $\phi_i^*=\langle a_i^\dagger \rangle$ and $\phi_j=\langle a_j \rangle$ are the SF order parameters. The resulting cluster Hamiltonian is then 
diagonalized self-consistently with respect to the superfluid order parameter $\phi_i$, while keeping all other parameters 
fixed. The ground state obtained in this way can be used to calculate the number of particles at each site as $\rho_i=\langle n_i \rangle $.

CMFT takes into account the non-local correlations which are otherwise overlooked in the single-site mean-field method and it is therefore 
more accurate. With proper implementation, results from CMFT can match fairly well with those obtained from other sophisticated 
methods like Quantum Monte Carlo, etc. but with significantly less computational efforts. Owing to these features, CMFT methods have been used 
extensively to successfully study a variety of problems in the past \cite{Penna, Hassan, Yamamoto, Macintosh, Dirk, MS1, MS2, MS4, AD}. In this work we use a four-site cluster as indicated by dashed lines in Fig.\ref{fig:schematic}, fix the value of $J$ as $1$ and scale all other parameters in units of $J$.

\subsection{Phase diagrams} \label{phase_dig}
\begin{figure}[!b]
 \centering
\includegraphics[width=0.46\textwidth]{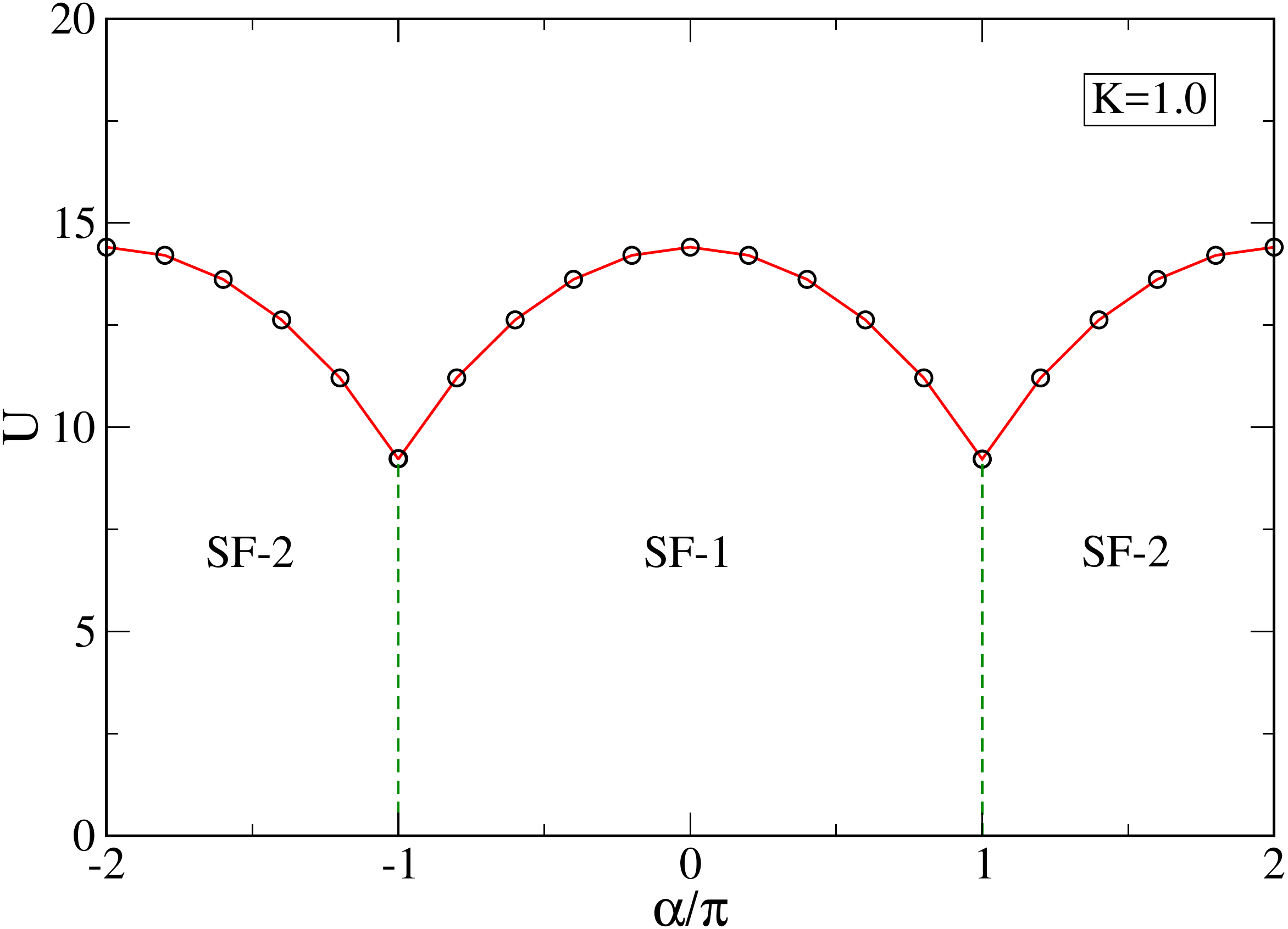}
  \caption{(Color online) Same as Fig. 3, but the results are obtained by using the CMFT approach. }
\label{fig:CMFT_phase_diagram_1}
\end{figure}

The phase diagram calculated using the CMFT method is shown in Fig.~\ref{fig:CMFT_phase_diagram_1}. To obtain it we first fix the value $K=1$  and choose a particular 
value of $\alpha(=n\pi)$. 
We then fix $U$ and vary $\mu$ to determine the $\phi_i$ self-consistently and
a vanishing value of $\phi_i$ along with an integer value of $\rho_i$ signifies the SF-MI transition. To obtain 
the critical point for the SF-MI transition, we increase the value of $U$ systematically until $\phi_i$ vanishes and 
$\rho_i$ becomes equal to 1, or in other words until the system enters the Mott-insulator phase with filling factor one. We repeat this procedure for several values of $\alpha$ varying from 
$-2\pi$ to $2\pi$ and the critical values of $U$ obtained in each case are marked by a black circle in the phase diagram  in 
Fig.~\ref{fig:CMFT_phase_diagram_1}. The continuous red line connecting the black circles then indicates the SF-MI 
phase boundary and by comparing these to Fig.~\ref{Fig:phase_ana}, one can clearly see that it matches the behaviour obtained using the Landau theory of phase transitions presented in Section \ref{Sec: Landau}. Numerically studying the cases for $J=1$ and $K \neq 1$ gives the  corresponding shifts in phase boundaries as well (not shown). 

\subsection{Chiral currents} \label{currents}

\begin{figure}
 \centering
 \vspace{0.5cm}
 \includegraphics[width=0.46\textwidth]{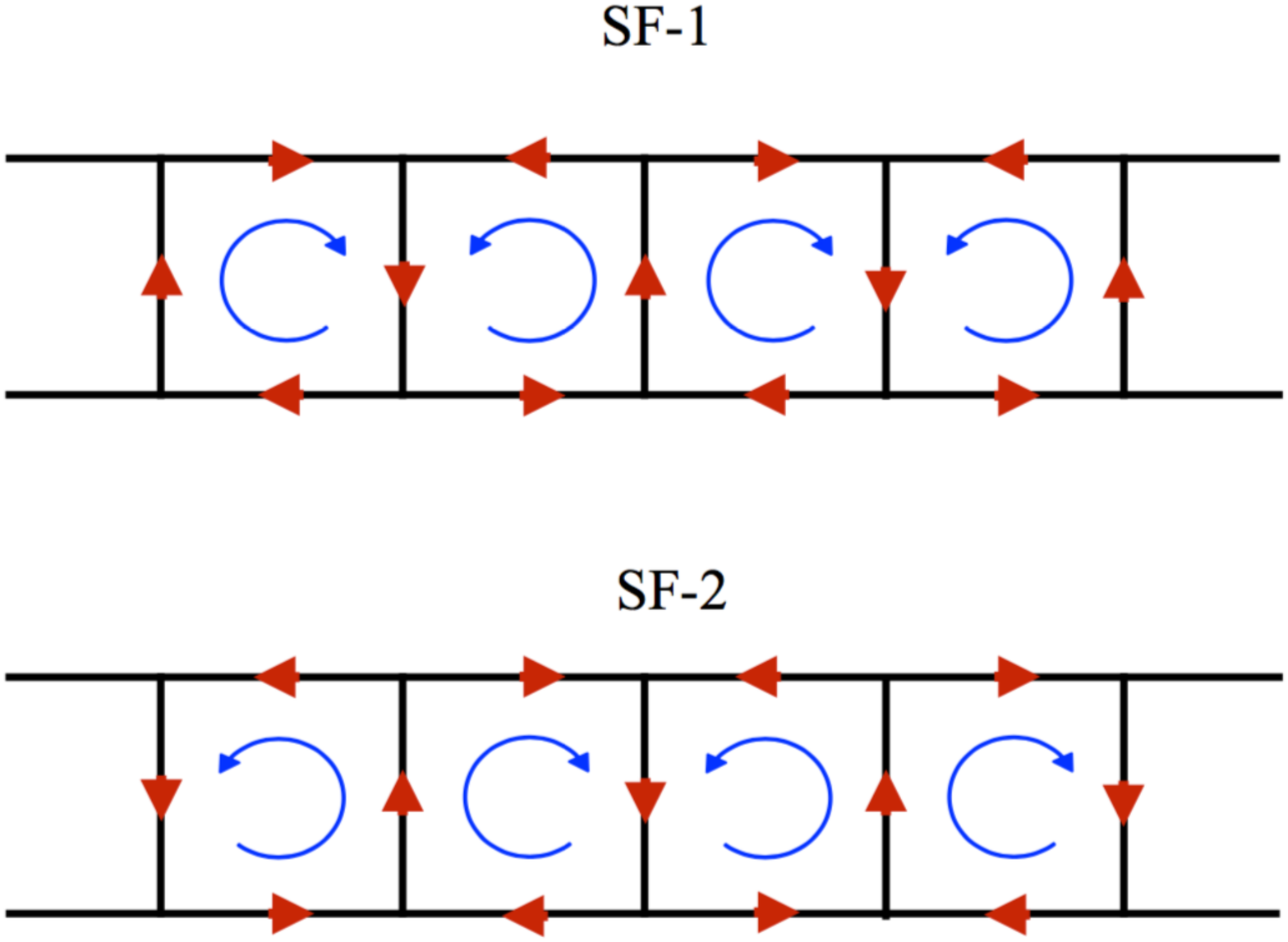}
 \caption{(Color online) Schematic of current patterns associated with the SF-1 and SF-2 phases. The red arrows denote the local currents given by equation (\ref{localcurrents}). The blue circular arrows denote the local staggered vortices/ anti vortices deduced from the local current pattern. The local currents possess opposite rotational directions for the two superfluid phases. }
\label{fig:current_schematic}
\end{figure}

We finally calculate the chiral currents in the system using CMFT, which will allow us to determine the overall flow pattern in the system.
The difference between the phases SF-1 and SF-2 can be characterized by their local current configurations and by their global chiral currents, the latter of which have the form
\begin{equation} 
 j_c=\sum_{l} \langle j_{l,b}^{||}-j_{l,a}^{||}\rangle,
\end{equation}
where the associated operators are
\begin{align}
 j_{l,a}^{||} &= iJ(e^{-i\alpha/2} a_{l+1}^\dagger a_l - e^{i\alpha/2} a_{l}^\dagger a_{l+1}), \nn\\
 j_{l,b}^{||} &= iJ(e^{i\alpha/2} b_{l+1}^\dagger b_l - e^{-i\alpha/2} b_{l}^\dagger b_{l+1}). \label{localcurrents}
\end{align}
\begin{figure}[b]
 \centering
 \vspace{0.5cm}
 \includegraphics[width=0.46\textwidth]{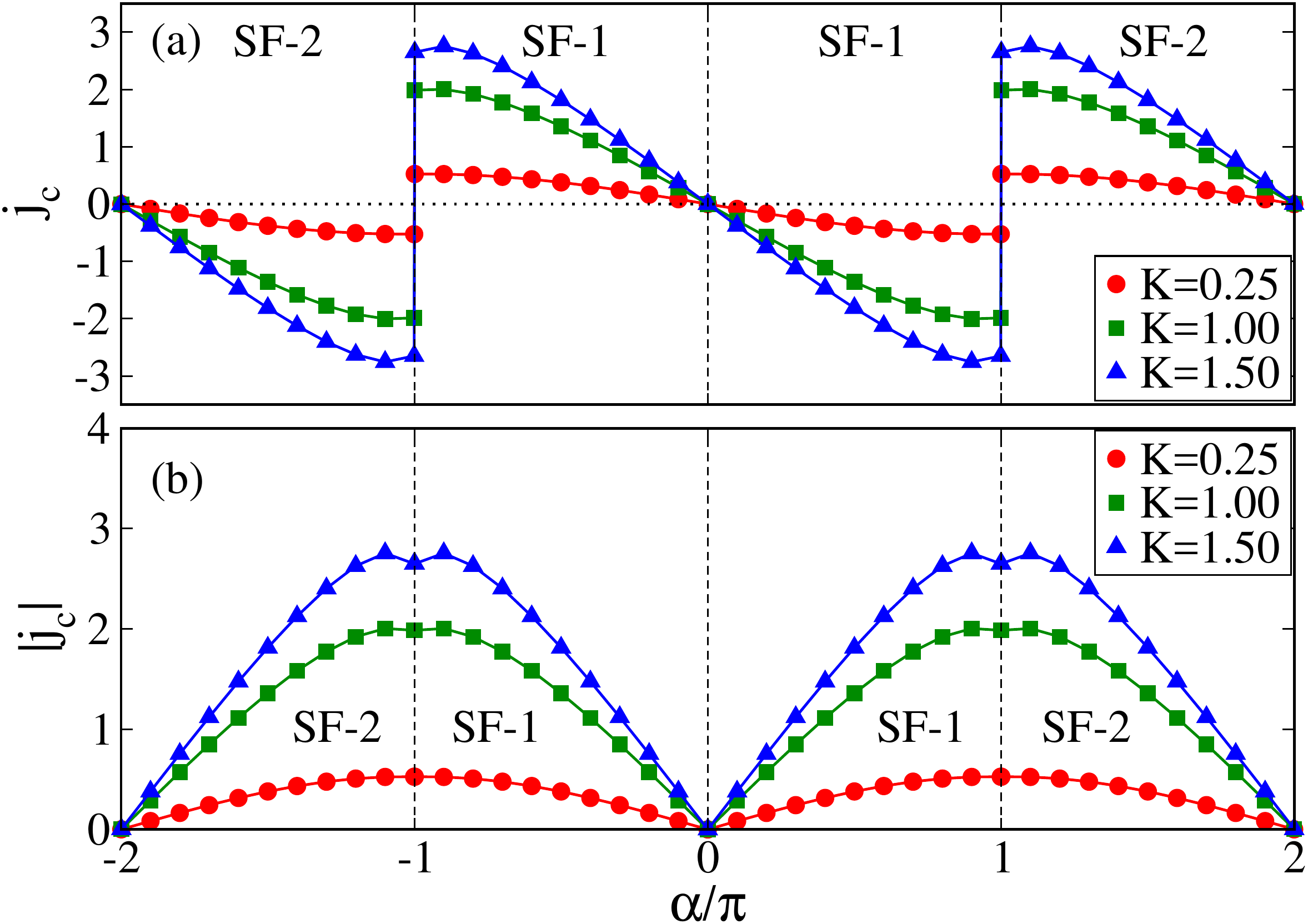}
 \caption{(Color online) Variation of $j_c$ (top panel) and $|j_c|$ (bottom panel) with $n$ for $J=1$ and $K=0.25, 1.0$ and $1.50$ and ($U,\mu$)
 $=(8.0,11.5$).}
\label{fig:CMFT_current}
\end{figure}
Here $l$ represent the site-index and for the numerical calculations we set the values of
on-site interaction to  $U=8$ and of the chemical potential to $\mu=11.5$, as for these parameters the system remains within the SF phase. The resulting chiral currents for different 
values of $K$ 
are shown in Fig.~\ref{fig:CMFT_current}. 
Two striking features are immediately obvious: (i) the sign of $j_c$ is reversed 
whenever the system makes a transition from the SF-1 to the SF-2 phase, while the sign of $\alpha$ is unchanged, and (ii) the slope of $|j_c|$ changes sign at the boundary between the two SF phases. The chiral currents for both SF-1 and SF-2 phases originate from the staggered currents going around each plaquette, and have opposite rotational directions in each phase. For the SF-1 phase, the value of chiral currents increases as a function of increasing magnetic flux $\alpha$, and local currents flowing around the plaquettes acquire a staggered (vortex-antivortex) configuration. At $\alpha=-\pi$ and $\pi$, the Hamiltonian becomes real and time-reversal invariant. Beyond these values, the staggered currents again break this symmetry, now with a reversal of the direction of the local currents around each plaquette, resulting in opposite chiral currents and a transition to SF-2 phase with a staggered (anti-vortex, vortex) current distribution. The flow of currents for both superfluid phases is schematically shown in Fig.~\ref{fig:current_schematic}.  Although the value of $\mu$ is fixed to $11.5$ for the chiral current calculations, we have checked and found similar results for other values of 
$\mu$ as well, as long as the system is in the superfluid phase. The only change is in the absolute value of $j_c$.

\section{Summary and outlook}\label{Sec: Summary}
We have examined the Bose Hubbard model in the presence of a staggered magnetic flux on a two-legged ladder configuration. We have shown that such a system possesses an interesting phase diagram, which is strongly influenced by the magnetic flux. The presence of alternating flux in the system leads to the appearance of two distinct superfluid phases, which are different to the ones observed in the standard two-leg Bose Hubbard model with uniform flux. We have performed numerical cluster mean field studies to confirm these analytically obtained phases.  
We believe that the model we have considered serves as an example for understanding the fundamental properties of lattices gases coupled to more complicated gauge fields, and can, in particular, stimulate experimental work on two-leg ladder bosonic systems in presence of staggered gauge fields. 

\section*{Acknowledgements}  This work was supported by the Okinawa Institute of Science and Technology Graduate University, Okinawa, Japan. 
The computational simulations were carried out using computing
facilities of Param-Ishan at Indian Institute of Technology,
Guwahati, India and Sango HPC facility at Okinawa Institute of
Science and Technology Graduate University, Okinawa,
Japan. M.S. acknowledges DST-SERB, India for the financial 
support through Project No. PDF/2016/000569. T.M. acknowledges Indian
Institute of Technology, Guwahati, India for the start-up grant
and DST-SERB, India for the financial support through Project
No. ECR/2017/001069.

\end{document}